\documentclass[twocolumn,showpacs,preprintnumbers,amsmath,amssymb]{revtex4}

\usepackage{graphicx}
\usepackage{dcolumn}
\usepackage{bm}
\usepackage{mathdots}

\begin{document}


\title{Simulating lattice gauge theories on a quantum computer}

\author{Tim Byrnes}

\email{tbyrnes@nii.ac.jp}
\affiliation{
National Institute of Informatics, 2-1-2 Hitotsubashi, Chiyoda-ku, Tokyo 101-8430, Japan}

\author{Yoshihisa Yamamoto}
\affiliation{E. L. Ginzton Laboratory, Stanford University, Stanford, CA 94305}
\affiliation{
National Institute of Informatics, 2-1-2 Hitotsubashi, Chiyoda-ku, Tokyo 101-8430, Japan}

\date{\today}

\begin{abstract}
We examine the problem of simulating lattice gauge theories on a universal quantum computer. 
The basic strategy of our approach is to transcribe lattice gauge theories in the Hamiltonian formulation
into a Hamiltonian involving only Pauli spin operators such that the simulation can be performed
on a quantum computer using only one and two qubit manipulations. We examine three models, the U(1), SU(2), and SU(3) 
lattice gauge theories which are transcribed into a spin Hamiltonian up to a cutoff in the Hilbert space of the
gauge fields on the lattice. 
The number of qubits required for storing a 
particular state is found to have a linear dependence with the total number of lattice sites.
The number of qubit operations required for performing the time evolution corresponding to the 
Hamiltonian is found to be between a linear to quadratic function of the number of lattice sites, depending 
on the arrangement of qubits in the quantum computer. 
We remark that our results may also be easily generalized to higher SU($N$) gauge theories.
\end{abstract}



\maketitle

\section{\label{sec:level1}Introduction}

The efficient simulation of quantum many-body problems is today of vital importance
to many fields of physics, stretching from condensed matter physics, atomic physics, molecular and optical physics, 
nuclear physics, quantum 
chemistry, and high-energy physics.  Despite advances in various numerical 
techniques, a truly reliable, accurate, and efficient method for extracting physical 
quantities for such problems does not exist in general.  For example, quantum Monte Carlo methods suffer
from the ``minus-sign problem,'' and the more recent Density Matrix Renormalization Group (DMRG)
method possesses difficulties in dimensions greater than one \cite{schollwock05}. The impact of a method that is free from 
such difficulties would undoubtedly offer new insights into the physics of quantum many-body systems. 

Quantum computers are hoped to offer an alternative to such methods. The exponential 
parallelism that is employed by quantum computers is believed to overcome the problems 
associated with the exponential explosion of Hilbert space size inherent to quantum many-body problems. It has been 
shown that an efficient simulation is indeed possible for a large class of spin, fermionic, and bosonic 
Hamiltonians \cite{ortiz01,somma03}. For example, fermionic Hamiltonians are implemented via an 
algebraic mapping of operators via the Jordan-Wigner transformation. This allows fermion operators appearing
in the Hamiltonian to be written entirely in terms of spin operators.  As discussed in Ref. \cite{ortiz01}, 
this approach is free of ``minus-sign problem'' that is present in Monte Carlo methods. 
The simulation is then performed 
by evolving the system according to the time evolution operator by a series of gates which evolve the system
according to the system Hamiltonian. A phase estimation algorithm using the quantum Fourier transform or a similar
algorithm is then performed to calculate the eigenenergies of a given system \cite{master03}. 
We assume implicitly here the standard model of quantum computation with two-level 
spin qubits and unitary operations. 
For bosonic operators, there is no exact algebraic mapping of operators into spin operators, 
as can be simply understood \cite{batista04} by the disparate sizes of the Hilbert spaces -- a single boson has
an infinite dimensional Hilbert space, whereas a spin has only two dimensions.  However, 
it is still possible to map the bosonic Hamiltonian onto a spin Hamiltonian via a transcription of the 
operator mappings \cite{somma03}. Clearly due to the mismatch in the system dimensions of the bosons and 
spins, there must be some truncation in the Hilbert space. However, this truncation can have little or no
effect in terms of the Hamiltonian if performed in the correct way. For example, Hamiltonians that
conserve particle number have a maximum of the total number of particles on a given site, and therefore has an 
effective cutoff equal to the number of particles. 

One problem that has received little attention is the problem of the quantum simulation of 
lattice gauge theories \cite{buchler05,osterloh05}.
Arguably this is the most computationally intensive quantum many-body problem of all, due to the large numbers
of degrees of freedom per site and the necessity of simulating in three spatial dimensions. Since the 
first Monte Carlo simulations over 25 years ago \cite{creutz79}, the lattice gauge theory program has 
steadily advanced with increases in computer speed and improvements in simulation techniques.
For the first time in the last few years it has become possible to perform a realistic simulation full QCD 
(i.e. with fermions) without quenching \cite{davies04}. However, should a
method become available that could give accurate results while being computationally
less expensive, this could offer a valuable alternative to the methods
being employed today. 

In this paper we study the problem of whether it is possible to implement lattice gauge theories
on a universal quantum computer efficiently. Before commencing we must make several choices in how
we formulate the problem. First, we assume that a quantum computer with individually addressable qubits 
and controllable local (nearest neighbor) two-qubit interactions 
is available. In particular, we assume that our quantum computer consists of two-level qubits,
so that it is operated on by Pauli spin operators. 
This is simply for convenience as most of the literature is formulated
in this way. Second, we choose to concentrate on pure gauge field theories alone, as an extension
to include fermions should involve only a small modification to the results presented here. For example, 
fermion fields may be implemented in a similar way to that described in Ref. \cite{ortiz01}. 
The third regards the formulation of the lattice gauge theory. To take
advantage of the quantum parallelism of the quantum computer we choose a Hamiltonian formulation 
of lattice gauge theory, in contrast to the Euclidean formulation that is more commonplace in present-day
simulations. The basic strategy is then to rewrite the operators appearing in the lattice gauge 
Hamiltonian into terms of spin operators. As is the case for bosons, no algebraic mapping of operators to 
spin operators is available for gauge operators due to different Hilbert space sizes. Analogously to boson
operators, we will show that it is possible to preserve the operator mappings of the gauge operators.
After this is done, one can implement the methods described in Refs. 
\cite{lloyd96,somma02,ortiz01} to perform the evolution. The Hamiltonian formulation has also the 
advantage that the lattice dimension for a simulation of real-world QCD (for example) is three rather 
than four for the Euclidean formulation, which reduces the space and time requirements for the calculation. 

We consider three models in this paper. First we consider compact U(1) lattice gauge theory. Although the 
transcription is particularly simple in this case, this will serve to introduce the basic strategy that
is employed. We then move onto the next most complicated case of SU(2) lattice gauge theory. This contains
all the elements that are required in order to transcribe any Hamiltonian for SU($N$). We finally consider 
the transcription of the SU(3) Hamiltonian, which is the closest relation to QCD. Although 
it is crucial that the time evolution operator can be implemented efficiently, it is also important that
other steps in the simulation can be carried out efficiently. To show that this is possible, in Sec. 
\ref{sec:init} we discuss methods for initializing the qubit states, suitable for extracting 
quantities such as the eigenenergies and expectation values of the low-lying eigenstates. This is 
again carried out for the U(1), SU(2), and SU(3) lattice gauge theories. 

This paper is organized as follows. In Sec. \ref{sec:prelim} we give a brief review of the 
simulation of spin Hamiltonians (Sec. \ref{sec:simspin}) and lattice gauge theory (Sec. \ref{sec:lgt}). 
In Sec. \ref{sec:form}, we transcribe the compact U(1) lattice gauge theory (Sec. \ref{sec:u1}), 
SU(2) lattice gauge theory (Sec. \ref{sec:su2}), and finally the SU(3) lattice gauge theory (Sec. \ref{sec:su3}) 
Hamiltonians. Sec. \ref{sec:init} 
discusses the initialization of qubits and Sec. \ref{sec:meas} discusses how observables may be extracted. Sec. \ref{sec:conc}
gives a summary of our findings and conclusions.

\section{\label{sec:prelim}Preliminaries}

\subsection{\label{sec:simspin}Simulating spin Hamiltonians}

We now give a brief review of the method of simulating spin Hamiltonians following Refs.  \cite{ortiz01,somma02}. 
The various methods of extracting eigenstates, eigenenergies, and expectation values for a given Hamiltonian 
in these references 
rely upon the ability of performing the time evolution corresponding to $ U(t) = \exp (-i H t ) $ where $ H $ is 
the Hamiltonian under investigation  \cite{lloyd96,master03,abrams99,ortiz01,somma02}. We cannot directly perform the 
evolution $ U(t) $ since we assume only one and two-qubit manipulations are available on our quantum computer. We 
therefore decompose the evolution operator using the Trotter formula \cite{lloyd96}:
\begin{equation}
e^{-iHt} = \lim_{m \rightarrow \infty} \left( e^{-i H_1 t/m} e^{-i H_2 t/m} \dots e^{-i H_{N_H} t/m} \right)^m ,
\label{trotter}
\end{equation}
where the Hamiltonian is a sum of $ N_H $ terms $ H = \sum_{i=1}^{N_H} H_i $. Assuming that the Hamiltonian 
under consideration here is entirely composed of Pauli spin operators, 
each term on the RHS of (\ref{trotter}) will consist of the exponential of a product of a number
of spin operators
\begin{equation}
U_i (t) = \exp ( -i \gamma_i \prod_j \sigma^{\alpha_j}_j ) ,
\label{trotterterm}
\end{equation}
where $ \gamma_i $ is a constant involving coupling constants and the time of evolution, $ j $ is an operator label (e.g. labeling the site number etc.), and 
$ \alpha_j = x, y, z $.  Operators such as
(\ref{trotterterm}) may be decomposed entirely into
one and two-qubit terms. Denote single qubit manipulations by
\begin{equation}
R_j (\theta,\bm{n}) = e^{i (\theta \bm{\sigma}_j \cdot \bm{n} + \delta )} \label{onequbit} ,
\end{equation}
where $ \delta $ is a global phase shift. Denote two-qubit manipulations by $ R_{jk}(\omega) $. For example, a
two-qubit manipulation that is available may be
\begin{equation}
R_{jk}(\omega) = e^{-i \omega \sigma^z_j \sigma^z_k} \label{twoqubit}  .
\end{equation}
The precise form of the two-qubit interaction is not important, so long that one
is present. Terms that involve only one spin operator in the 
exponent in (\ref{trotterterm}) may be directly constructed from (\ref{onequbit}).  Terms involving
a product of two or more spin operators in the exponent of (\ref{trotterterm}) can be written by an 
appropriate combination of one and two-qubit operations. For example, say we need to construct
the operator $ e^{-i \sigma^x_1 \sigma^y_2} $. This may be decomposed into the product
\begin{eqnarray*}
e^{-i \sigma^x_1 \sigma^y_2} = R_2^\dagger (\pi/4,\bm{x}) R_1^\dagger (\pi/4,\bm{y}) R_{12}(1) R_1 (\pi/4,\bm{y}) R_2 (\pi/4,\bm{x}),
\end{eqnarray*}
where we assumed the two qubit-manipulation (\ref{twoqubit}) and used the 
identity $ U^\dagger e^M U = e^{U^\dagger M U} $, where $ U $ is an unitary operator 
and $ M$ is an arbitrary operator. 
Longer products of spin operators may be constructed by successive products of two-qubit 
operators \cite{ortiz01}. 

The simulation then proceeds in the following way. The initial state of the qubits is first prepared, 
the precise form of which depends on the quantity that is being calculated. This initial 
state is then evolved according to the particular algorithm to be performed, which implements
the time evolution operator $ U(t) = e^{-iHt} $. The time evolution is performed according to the method described 
above. Finally a measurement is performed on the system to extract the desired results. For example, in order
to measure the eigenvalue spectrum, the qubits are prepared in a state with a non-zero
overlap with the eigenstate of interest. This is typically the ground or low-lying eigenstates of the 
system. A suitable initial state is for example the mean-field ground state of the Hamiltonian. 
This initial state is then evolved forwards using the time evolution operator $ U(t) $, using
a phase estimation algorithm  \cite{abrams99}. The estimated phases then 
recover the eigenspectrum of the Hamiltonian. 
Further details of the extraction of observables may be found in Refs.
\cite{lloyd96,master03,abrams99,ortiz01,somma02}.

\subsection{\label{sec:lgt}Lattice gauge theory}

\begin{figure}
\scalebox{0.5}{\includegraphics{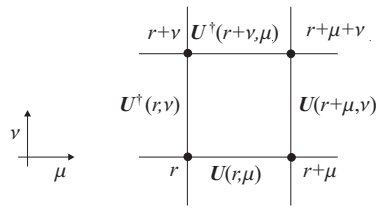}}
\caption{\label{fig:plaquette} An elementary plaquette. Each link is labeled by the coordinate $ r $ of 
one of the sites that it connects, and the direction of the link $ \mu, \nu $.}
\end{figure}

Let us define SU($N$) lattice gauge theory in the Hamiltonian formulation \cite{kogut75} (an excellent review of 
lattice gauge theory in the Hamiltonian formulation may be found in Ref. \cite{kogut79}). 
Assume a $d$-dimensional lattice.  Label the sites of the lattice
by a coordinate $ r $. Each site has $ 2d$ links emanating from it. Label each of these 
links according to the site it is attached to, and the direction that the link points (see Fig. 
\ref{fig:plaquette} for the labeling conventions). 
On each link of the lattice define an $ N^2 - 1 $ component operator
$ \bm{A}^\alpha $, with canonically conjugate operators $ \bm{E}^\alpha$. These 
are the gauge field operators of the lattice gauge theory. Define an
$N \times N $ unitary matrix involving the gauge fields
\begin{equation}
\bm{U} (r,\mu) \equiv \exp( i \tfrac{1}{2} ga \sum_{\alpha = 1}^{N^2 -1} \tau^\alpha \bm{A}^\alpha (r,\mu) ) 
\label{generalu}
\end{equation}
where $ \tau^\alpha $ ($\alpha = 1, \dots, N^2 - 1 $) are the fundamental generators of SU($N$), 
$g$ is the coupling constant, and $a$ is the lattice spacing. The generators of SU($N$) obey
\begin{equation}
\mbox{Tr} ( \tau^\alpha \tau^\beta ) = \tfrac{1}{2} \delta^{\alpha \beta} 
\end{equation}
and
\begin{equation}
[ \tau^\alpha, \tau^\beta ] = i f^{\alpha \beta \gamma} \tau^\gamma,
\end{equation}
where $ \delta^{\alpha \beta}  $ is the Kronecker delta and $ f^{\alpha \beta \gamma} $ are the structure 
constants of the group. The following commutation relations then follow
\begin{eqnarray}
\left[ \bm{E}^\alpha, \bm{U}_{ij} \right]  & = &  \tfrac{1}{2}  \sum_{k=1}^N \tau_{ik}^\alpha \bm{U}_{kj} 
\label{generalcom1} \\
\left[ \bm{E}^\alpha, \bm{U}_{ij}^\dagger \right]  & =  & - \tfrac{1}{2} \sum_{k=1}^N  \bm{U}_{ik}^\dagger \tau_{kj}^\alpha 
\label{generalcom2} \\
\left[ \bm{E}^\alpha, \bm{E}^\beta \right] & = & i f^{\alpha \beta \gamma} \bm{E}^\gamma
\label{generalcom3}
\end{eqnarray}
where the indices $i,j$ on $ \bm{U}_{ij} $ refer to the matrix elements of the $ N \times N $ matrix. 
We denote all quantities which contain the operators $ \bm{E}^\alpha$ or $\bm{A}^\alpha $ in bold font. 
The SU($N$) pure gauge Hamiltonian is written
\begin{eqnarray}
\bm{H} & = & \frac{g^2}{2a} \Big[ \sum_{r, \mu} \sum_\alpha \left( \bm{E}^\alpha (r,\mu) \right)^2  \nonumber \\
& & + x \sum_{p \in \mbox{\tiny \{plaquettes\}}} \left( 2N - (\bm{Z}(p) + \bm{Z}^\dagger(p)) \right) \Big]
\label{generalham}
\end{eqnarray}
where 
\begin{equation}
\bm{Z}(p) =  \mbox{Tr} \left[ \bm{U}  (r,\mu) \bm{U} (r+ \mu,\nu ) \bm{U}^\dagger (r+\nu,\mu) \bm{U}^\dagger (r,\nu) \right]
\label{generalplaq}
\end{equation}
and $ x = \frac{2}{g^4} $. The trace in the $ \bm{Z}(p) $ operator is taken after matrix multiplication of the four 
$ \bm{U} $ matrices. The first summation in the Hamiltonian (\ref{generalham}) 
sums over all links in the lattice, and the summation involving $ \bm{Z}(p) $ operator 
sums over all plaquettes in the lattice. The 
arrangement of links in a plaquette operator is shown in Fig. \ref{fig:plaquette}. 
The overall constant of $ g^2/2a $ and the constant term $ 2 N $ in the plaquette summation 
merely rescales and shifts the energies respectively and will be omitted for the rest of this
paper.

\section{\label{sec:form}Formulating lattice gauge theories via spin operators}

\subsection{\label{sec:u1}Hamiltonian for U(1) lattice gauge theory}

We first start with one of the simplest lattice gauge theories: compact U(1) lattice gauge theory (compact quantum 
electrodynamics with no fermions).  The transcription of the operators is particularly simple in 
this case and will therefore serve to introduce the basic method that will be employed for the SU(2) and SU(3)
cases.  The U(1) Hamiltonian is \cite{kogut75}
\begin{equation}
\bm{H}_{\mbox{\tiny U(1)}} = \sum_{r,\mu}\bm{E}^2 (r,\mu)  -
x \sum_{p \in \mbox{\tiny \{plaquettes\}}} \left( \bm{Z}(p) + \bm{Z}^\dagger(p) \right) 
\end{equation}
where
\begin{equation}
\bm{Z}(p) = e^{  i \bm{\theta}(r,\mu) }  e^{  i \bm{\theta}(r+ \mu,\nu ) } 
 e^{  -i \bm{\theta}(r+\nu,\mu) } e^{ -  i \bm{\theta}(r,\nu) }
\label{u1zp}
\end{equation}
where we have defined $ \bm{\theta} = ga \bm{A} $ according to standard notation. 
The generator of U(1) is a scalar and therefore all the $ \bm{U} = e^{  i \theta} $ commute with each other. 
The $ \bm{U} $ are $ 1 \times 1 $ matrices.  The $ \bm{E} $ obey the commutation
relations
\begin{eqnarray}
\left[ \bm{E}, \bm{U} \right]  & = &  \bm{U} \\
\left[ \bm{E}, \bm{U}^\dagger \right]  & =  & - \bm{U}^\dagger .
\end{eqnarray}
The Hilbert space for the Hamiltonian may be generated by successive applications of $ \bm{Z}(p) + \bm{Z}^\dagger(p) $
on the state $ | 0 \rangle $ which is defined to be an eigenstate of the $ \bm{E} $ operator satisfying 
\begin{equation}
\bm{E}(r,\mu) | 0 \rangle = 0
\label{u1strongcoup}
\end{equation}
for all $ r$ and $\mu $  \cite{kogut75}. 

\begin{figure}
\scalebox{0.4}{\includegraphics{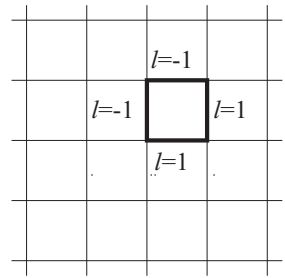}}
\caption{\label{fig:u1} A single plaquette excitation for U(1) lattice gauge theory. All unlabeled links have
$ l = 0 $. }
\end{figure}

Let us first consider the Hilbert space of a single link. 
A link where $ \bm{U} $ has been applied $ l $ times is an eigenstate of $ \bm{E}^2 $ with eigenvalue $l^2 $:
\begin{equation}
\bm{E}^2 ( \bm{U} )^l | 0 \rangle = l^2 ( \bm{U} )^l | 0 \rangle .
\end{equation}
We can therefore identify the orthogonal states
\begin{equation}
| l \rangle = ( \bm{U} )^l | 0 \rangle
\end{equation}
which satisfy
\begin{eqnarray}
\bm{E}^2 | l \rangle   & = & l^2 | l \rangle \\
\bm{U} | l \rangle   =  | l + 1 \rangle  \hspace{6mm} & & \hspace{4mm} 
\bm{U}^\dagger | l \rangle  =  | l - 1 \rangle. \label{raisinglowering}
\end{eqnarray}
The plaquette operator then shifts the $ | l \rangle $ eigenstates on links arranged on all squares of the lattice
(see Fig. \ref{fig:u1}).

This space of states may be easily rewritten in the spin language in the following way. On a given link,
a quantum register of $ D $ qubits keeps track of the $ | l \rangle $ state that the link is in. Denote the 
state that the register is in by $ | l \rangle_{\mbox{\tiny reg}} $, where the ``reg'' serves as a reminder that this is a state of 
the register. Clearly, since
$ l =\{- \infty, \dots,  \infty \} $ and $ D $ qubits can only keep track of a maximum of $ 2^D $ values $ l $, 
some truncation is involved. This truncation is performed such that 
$ l =\{ -l_{\mbox{\tiny max}}, \dots, l_{\mbox{\tiny max}} \} $.
$ D $ qubits will then give an $ l_{\mbox{\tiny max}} $ of $ (2^D -1)/2 $. The impact of this truncation can be
handled by analyzing the convergence of observables (e.g. the energy spectrum) with 
$ l_{\mbox{\tiny max}} $ \footnote{Past studies
suggest that a small value of $ l_{\mbox{\tiny max}} $ is usually sufficient for retaining a high degree of numerical
precision \cite{irving83,byrnes02}. For example, in Ref. \cite{byrnes02} which examined one dimensional quantum
electrodynamics, an $ l_{\mbox{\tiny max}} \approx 5 $ was sufficient to retain an accuracy to machine precision.}.
Now define operators that act on these registers
\begin{eqnarray}
E^2 | l \rangle_{\mbox{\tiny reg}} & = & l^2 | l \rangle_{\mbox{\tiny reg}} \label{e2operatoru1} \\
L^\pm | l \rangle_{\mbox{\tiny reg}} & = & | l \pm 1 \rangle_{\mbox{\tiny reg}} \\
L^+ | l_{\mbox{\tiny max}} \rangle_{\mbox{\tiny reg}} & = & 0 \\
L^- | -l_{\mbox{\tiny max}} \rangle_{\mbox{\tiny reg}} & = & 0
\label{voperatoru1}
\end{eqnarray}
A simple implementation of these operators is given in Appendix \ref{sec:apexample}.
It is then a simple
matter to rewrite the U(1) Hamiltonian as
\begin{equation}
H_{\mbox{\tiny U(1)}} = \sum_{r,\mu} E^2 (r,\mu)  -
x \sum_{p \in \mbox{\tiny \{plaquettes\}}} \left( Z(p) + Z^\dagger (p) \right) 
\label{u1spinhamil}
\end{equation}
where
\begin{eqnarray*}
Z(p) & = & L^+ (r,\mu) L^+ (r+ \mu,\nu ) L^-  (r+\nu,\mu) L^- (r,\nu) , \\
Z^\dagger  (p) & = & L^- (r,\mu) L^- (r+ \mu,\nu ) L^+  (r+\nu,\mu) L^+ (r,\nu) ,
\end{eqnarray*}
A similar Hamiltonian was given in Refs.  \cite{chandrasekharan97,batista04}. 

\begin{figure}
\scalebox{0.8}{\includegraphics{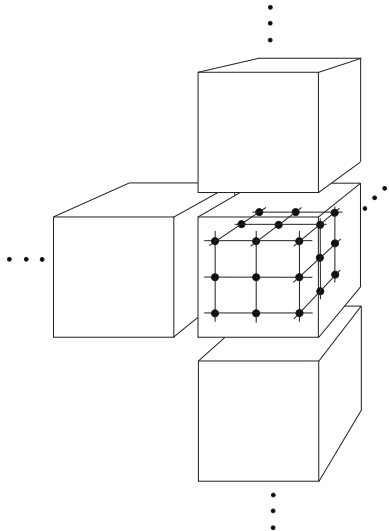}}
\caption{\label{fig:qubitarrangement} An optimal arrangement of qubits for lattice gauge theories. The qubits associated
with a particular link are shown in boxes. Registers
keeping track of the state of a particular link are spatially arranged in the same way as the links of 
the original lattice.}
\end{figure}

For a total 
of $ M $ lattice sites, this requires a total of $ \sim M d D $ qubits and is therefore linear
in space requirements. Let us estimate the dependence of the total number of operations required for 
a particular time evolution $ U(t) $ with respect to the number of lattice sites $M$. 
By a judicious arrangement of qubits, it is clear that the $ M $ dependence can be made linear. 
In Fig. \ref{fig:qubitarrangement}, the qubits associated with an $ | l \rangle $ register for a particular 
link are grouped spatially in the same location, with the same units repeating in the same configuration 
as the links on the original lattice. As each term in the Hamiltonian (\ref{u1spinhamil}) only acts
between nearest neighbors, there is no dependence on $ M $ for each term. The 
total number of operations is therefore proportional to the number of terms in the Hamiltonian, which 
is $ \propto M d $. A less ideal situation will have a qubit arrangement which involves manipulations
between qubits which are spatially separated by distances of the order of the number of qubits 
required for the simulation. In the worst case, every term in the Hamiltonian will involve qubit operations
between very distant qubits. The number of operations, i.e. the sequence of swap operations, 
required for executing such a long-range term is of the order of 
the total number of qubits $ \sim M d D $. Therefore, the total number of 
operations will grow $ \propto M^2 d^2 D $. This is quadratic with the lattice size. The best 
case and the worst case scenarios presented here are both probably rather unrealistic; we expect that the 
true dependence will fall somewhere in between.

\subsection{\label{sec:su2}Hamiltonian for SU(2) lattice gauge theory}

We now examine SU(2) lattice gauge theory. The Hamiltonian
is \cite{kogut75}
\begin{equation}
\bm{H}_{\mbox{\tiny SU(2)}} = \sum_{r, \mu} \sum_\alpha \left( \bm{E}^\alpha (r,\mu) \right)^2  -
2x \sum_{p \in \mbox{\tiny \{plaquettes\}}} \bm{Z}(p) 
\label{hamsu2}
\end{equation}
where $ \bm{Z}(p) $ was defined in (\ref{generalplaq}). A simplification occurred in the above
Hamiltonian due to $ \bm{Z}^\dagger (p) = \bm{Z}(p)$ for SU(2). The $ \bm{U}$ operators appearing
in the plaquette operator $ \bm{Z}(p) $ are now $ 2 \times 2 $ matrices. The generators $ \tau^\alpha $
in (\ref{generalu}) are the $ 2 \times 2 $ Pauli matrices.  The commutation relations of the $ \bm{E}^\alpha$ 
with the $ \bm{U} $ obey relations (\ref{generalcom1}) - (\ref{generalcom3}). 

\begin{figure}
\scalebox{0.4}{\includegraphics{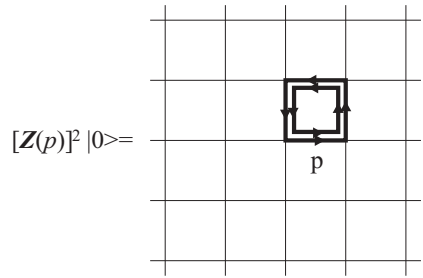}}
\caption{\label{fig:su2} A double plaquette excitation. All unlabeled links have
$ \bm{E}^\alpha (r,\mu) | 0 \rangle = 0 $. }
\end{figure}

The Hilbert space of the Hamiltonian is generated by successive applications of $ \bm{Z} (p) $ on all possible 
locations on the lattice starting from the state satisfying $ \bm{E}^\alpha (r,\mu) | 0 \rangle = 0 $ for all $ r$ and $\mu $. 
Consider a state such as that shown in Fig. \ref{fig:su2}, where two plaquette operators $ \bm{Z} (p) $ operate on the same 
plaquette. Concentrating on one of the four links, this creates terms such as
\begin{equation}
\bm{U}_{m_L m_R} \bm{U}_{n_L n_R} | 0 \rangle,
\label{twotermssu2}
\end{equation}
where there is no matrix multiplication between the $\bm{U}$'s as all indices have been contracted in the trace of each 
$ \bm{Z} (p) $ operator. We distinguish between the left and right matrix element labels using the 
subscripts $L $ and $R$ on the matrix element labels. Such terms may be simplified using the group theoretic identity \cite{swart63,kogut76}
\begin{eqnarray}
&& \hspace{-7mm} U_{m_L m_R}^{j'}(\phi) U_{n_L n_R}^{j}(\phi) =  \nonumber \\
&&  \hspace{-5mm} \sum_{J = | j - j' |}^{j + j'} 
\langle J, M_L | j,n_L; j',m_L \rangle \langle J,M_R |  j,n_R; j',m_R \rangle U^J_{M_L M_R} (\phi) ,\nonumber \\
\label{ucombsu2}
\end{eqnarray}
where
\begin{equation}
U_{m n}^j(\phi)  = \left[ \exp( i \sum_{\alpha = 1}^3 \tau^\alpha_j \phi^\alpha ) \right]_{m n} ,
\end{equation}
and $ \tau^\alpha_j $ is the generator for the irreducible representation of SU(2) corresponding to angular momentum $j$ ($ \tau^\alpha_j $
is a $ (2j+1) \times (2j+1) $ matrix).  The $ \langle J, M | j,n; j',m \rangle $ denote Clebsch-Gordan coefficients
for SU(2). We may now evaluate (\ref{twotermssu2}). Let us generalize this a little by taking
$ \bm{U}_{n_L n_R} \rightarrow \bm{U}_{n_L n_R}^{j} $. 
Since the $ \bm{U} $ operators appearing in $ \bm{Z}(p) $ are all in the $ j = 1/2 $ representation, 
this corresponds to $j' = 1/2 $ in (\ref{ucombsu2}) and thus there are only two terms in the summation: 
$ J = j \pm 1/2 $ (unless $ j = 0 $). It is also convenient to define  \cite{robson82}
\begin{equation}
| j, n_L, n_R  \rangle \equiv \sqrt{2j+1} \bm{U}_{n_L n_R}^j | 0 \rangle ,
\end{equation}
where the factor of $ \sqrt{2j+1} $ necessary so that the states are correctly normalized:
\begin{equation}
\langle j', m_L, m_R | j, n_L, n_R  \rangle = \delta_{j j'} \delta_{m_L n_L} \delta_{m_R n_R} .
\end{equation}
We then have
\begin{eqnarray}
& &   \bm{U}_{m_L m_R}^{1/2} | j, n_L, n_R \rangle = \nonumber \\
& & \hspace{-4mm} \sum_{J = | j - 1/2 |}^{j+1/2} 
\sqrt{\frac{2j+1}{2J+1}}
\langle J, M_L  |  j, n_L; \tfrac{1}{2}, m_L \rangle 
\langle J, M_R  |  j, n_R; \tfrac{1}{2}, m_R \rangle \nonumber \\
& & \hspace{10mm} \times  | J, M_L=m_L+n_L, M_R=m_R + n_R  \rangle . \label{uapplicationsu2}
\end{eqnarray}
The difference to the analogous result for the U(1) case (see Eqn. (\ref{raisinglowering})) is 
that there are now two terms and Clebsch-Gordan coefficients are present. 

Let us also find the effect of operating the first term in the Hamiltonian (\ref{hamsu2}) on the 
state $ | j, n_L, n_R  \rangle $. Using the 
commutation relation
\begin{eqnarray*}
\left[ \sum_\alpha (\bm{E}^\alpha)^2, \bm{U}_{n_L n_R}^j \right]  
& = &  \tfrac{1}{4} \sum_{\alpha=1}^3 \sum_m [( \tau^\alpha)^2]_{n_L m} \bm{U}_{m n_R}^j \\
& = & j(j+1) \bm{U}_{n_L n_R}^j ,
\end{eqnarray*}
we have
\begin{equation}
\sum_\alpha (\bm{E}^\alpha)^2 | j,n_L, n_R  \rangle = j(j+1) | j,n_L, n_R  \rangle .
\label{casimirsu2}
\end{equation}

Our problem is then to rewrite Eqns. (\ref{uapplicationsu2}) and (\ref{casimirsu2}) into a spin
language. A similar problem was examined by Bacon, Chuang, and Harrow  \cite{bacon04}, where an
algorithm was presented to perform the transformation 
\begin{eqnarray}
& &  | j,m \rangle_{\mbox{\tiny reg}}  | \tfrac{1}{2}, \Delta m \rangle_{\mbox{\tiny reg}}  \rightarrow  \nonumber \\
& & \sum_{J = |j - 1/2|}^{j+1/2} \langle J, M | j, m ;  \tfrac{1}{2}, \Delta m  \rangle | 
J, M = m + \Delta m \rangle_{\mbox{\tiny reg}} . \nonumber \\
\label{baconcircuit}
\end{eqnarray}
The Clebsch-Gordan coefficients $ \langle J, M |   j, m; \tfrac{1}{2}, \Delta m \rangle $ may be calculated using the
formulas presented in Table \ref{tab:su2coeffs}. A completely general way of calculating such 
coefficients for SU($N$) is presented in Refs.  \cite{biedenharn68,louck70}. The circuit of Bacon, Chuang, 
and Harrow may be rewritten in operator form defining the operators
\begin{table}
\caption{\label{tab:su2coeffs}Formulas for Clebsch-Gordan coefficients 
$ \langle J= j+ \Delta j, M= m_i +\Delta m|  j, m_i ;\tfrac{1}{2}, \Delta m \rangle $, where 
$ i = L, R $ is put in as appropriate.  The Wigner
operators corresponding to the shifts are also listed for comparison 
with Refs.  \protect\cite{biedenharn68,louck70}.}
\begin{ruledtabular}
\begin{tabular}{ccccc}
Coefficient & Wigner & $ \Delta j $ & $ \Delta m $ & Formula \\
& operator \\
\hline
$ c_{11}^i$  & 
$ \left\langle
\begin{array}{c}
 1  \\
1 \ 0 \\
1 \end{array}
\right\rangle $
& 1/2 & 1/2 & $ \sqrt{\frac{j+m_i+1}{2j+1}} $ \\
$ c_{12}^i$  & 
$ \left\langle
\begin{array}{c}
1  \\
1 \ 0 \\
0  \end{array}
\right\rangle $
& 1/2 & -1/2 & $ \sqrt{\frac{j-m_i+1}{2j+1}} $ \\
$ c_{21}^i$  & 
$ \left\langle
\begin{array}{c}
0  \\
1 \ 0 \\
1 \end{array}
\right\rangle $
& -1/2 & 1/2 & $  -\sqrt{\frac{j-m_i}{2j+1}} $ \\
$ c_{22}^i$  & 
$ \left\langle
\begin{array}{c}
0  \\
1 \ 0 \\
0  \end{array}
\right\rangle $
& -1/2 & -1/2 & $ \sqrt{\frac{j+m_i}{2j+1}} $ \\
\end{tabular}
\end{ruledtabular}
\end{table}
\begin{eqnarray}
J^\pm | j\rangle_{\mbox{\tiny reg}} & = & | j \pm \tfrac{1}{2} \rangle_{\mbox{\tiny reg}} \\
M^\pm | m  \rangle_{\mbox{\tiny reg}} & = & | m \pm \tfrac{1}{2} \rangle_{\mbox{\tiny reg}} .
\end{eqnarray}
For the two cases $ \Delta m = +\tfrac{1}{2}$ and $  \Delta m = -\tfrac{1}{2} $, we have 
\begin{eqnarray}
V(\Delta m=1/2) & = & M^+  (  J^+ \hat{c}_{11} +  J^- \hat{c}_{21}) \\ 
V(\Delta m=-1/2) & = & M^-  (  J^+ \hat{c}_{12} +  J^- \hat{c}_{22})  ,
\end{eqnarray}
where the $ \hat{c}_{mn} $ are diagonal operator versions of the $ c_{mn} $ coefficients that extract the 
coefficients from $ | j,m  \rangle_{\mbox{\tiny reg}} $ using the formulas presented in Table \ref{tab:su2coeffs}.  
An example of how such an operator may be constructed is shown in Appendix \ref{sec:apexample}. 
One may verify that this is equivalent to the transformation (\ref{baconcircuit}) by applying the $ V(\Delta m) $ operators
to the state $ | j,m \rangle_{\mbox{\tiny reg}} $.

It is now clear how to write an operation corresponding to (\ref{uapplicationsu2}). For each link, we require three
registers corresponding to the state $ | j,m_L,m_R  \rangle $. Denote this by $ | j,m_L,m_R \rangle_{\mbox{\tiny reg}} $. 
As for the U(1) case, a maximum cutoff must be imposed on these registers to keep the number of qubits finite. For
$ j = \{ 0, \tfrac{1}{2}, \dots, j_{\mbox{\tiny max}} \} $, the $ m_L $ and $ m_R $ registers will be in the range
$ \{ -j_{\mbox{\tiny max}}, -j_{\mbox{\tiny max}} + \tfrac{1}{2} ,\dots, j_{\mbox{\tiny max}} \} $.  

We may then write the operator corresponding to (\ref{uapplicationsu2}) as
\begin{equation}
\bm{U}_{mn}^{1/2} \leftrightarrow V_{mn} =
M_{m}^L M_{n}^R \left[  J^+ \hat{c}_{1 m}^L \hat{c}_{1 n}^R \hat{N}_1 +  J^- 
\hat{c}_{2 m}^L \hat{c}_{2 n}^R \hat{N}_2 \right]  ,
\end{equation}
where we defined
\begin{equation}
M_m^L | j,n_L,n_R   \rangle_{\mbox{\tiny reg}} = 
\left\{ 
\begin{array}{ll}
| j, n_L + \tfrac{1}{2} ,n_R   \rangle_{\mbox{\tiny reg}} & \hspace{1cm} m=1 \\
| j, n_L - \tfrac{1}{2},n_R   \rangle_{\mbox{\tiny reg}} & \hspace{1cm} m=2 
\end{array}
\right.
\end{equation}
and
\begin{equation}
M_m^R | j,n_L,n_R   \rangle_{\mbox{\tiny reg}} = 
\left\{ 
\begin{array}{ll}
| j, n_L ,n_R + \tfrac{1}{2}  \rangle_{\mbox{\tiny reg}} & \hspace{1cm} m=1 \\
| j, n_L,n_R  - \tfrac{1}{2}  \rangle_{\mbox{\tiny reg}} & \hspace{1cm} m=2 
\end{array}
\right. 
\end{equation}
and also
\begin{equation}
\label{normalsu2}
N_k = \left\{ 
\begin{array}{ll}
\sqrt{\frac{2j+1}{2j+2}} & \hspace{1cm} k=1 \\
\sqrt{\frac{2j+1}{2j}} & \hspace{1cm} k=2 
\end{array}
\right.  .
\end{equation}
The diagonal operators $ \hat{N}_k $ recover the normalization factors from the registers 
$ | j,n_L,n_R   \rangle_{\mbox{\tiny reg}} $ according to (\ref{normalsu2}).

We may now write the full Hamiltonian for SU(2) using these operators. 
It is first convenient to rewrite (\ref{generalplaq}) using the identity for SU(2) \cite{robson82}
\begin{equation}
\bm{U}^\dagger_{mn} = (-1)^{m - n} \bm{U}_{-m -n}, 
\label{su2identity}
\end{equation}
which gives
\begin{eqnarray}
\bm{Z}(p) & = & \sum_{m_1 m_2 m_3 m_4} (-1)^{m_1 - m_3} \bm{U}_{m_1 m_2} (r,\mu) 
 \nonumber  \\
& & \hspace{-5mm} \times \bm{U}_{m_2 m_3} (r+ \mu,\nu )
\bm{U}_{-m_3 -m_4}(r+\nu,\mu) \bm{U}_{-m_4 -m_1} (r,\nu) .
\nonumber \\
\end{eqnarray}
Let us also define the operator 
\begin{equation}
E^2 | j, m_L, m_R  \rangle_{\mbox{\tiny reg}} = j(j+1) | j, m_L, m_R \rangle_{\mbox{\tiny reg}}.
\end{equation}
We thus have
\begin{equation}
H_{\mbox{\tiny SU(2)}}  = \sum_{r, \mu} E^2 (r,\mu) 
-2x \sum_{p \in \mbox{\tiny \{plaquettes\}} } Z(p) 
\end{equation}
where
\begin{eqnarray}
Z(p) & = &
\sum_{m_1 m_2 m_3 m_4} (-1)^{m_1 - m_3} V_{m_1 m_2} (r,\mu) \nonumber \\
& & \times V_{m_2 m_3} (r+ \mu,\nu )   V_{-m_3 -m_4}(r+\nu,\mu) V_{-m_4 -m_1} (r,\nu) . \nonumber \\ 
\end{eqnarray}
This is the final result for SU(2).  The Hermiticity of this Hamiltonian is guaranteed from the property
\begin{equation}
V_{m n}^\dagger = (-1)^{m-n} V_{-m -n} .
\end{equation}

Let us now estimate the number of qubits and the number of operations required for this Hamiltonian. 
Three registers are required per lattice 
link and therefore the number of qubits is $ \sim 3 M D d $, where $ D $ is the number of qubits in a 
single register. This is linear with the number of sites $ M $. As for the U(1) case, with a judicious 
arrangement of qubits (see Fig. \ref{fig:qubitarrangement}), the local nature of the Hamiltonian 
may be preserved in the transcribed version of the Hamiltonian. This makes the total number 
of operations $ \propto M d $ as for the U(1) case. The proportionality constant will be larger in 
this case, as each lattice link requires a larger number of qubits. There is also the added complexity 
in calculating the Clebsch-Gordan coefficients $ \hat{c}_{ij} $, which may be performed in 
$ \mbox{poly} (j_{\mbox{\tiny max}})$ operations. 
Again the worst-case estimate may be obtained by assuming that each term in the Trotter expansion
is a product of operators with a length of the order of the number of qubits. We therefore obtain
the same result as the U(1) case, such that the total number of operations is $ \propto M^2 d^2 D $.

\subsection{\label{sec:su3}Hamiltonian for SU(3) lattice gauge theory}

The last case we consider is SU(3) lattice gauge theory, which is the theory of principal
interest as it is closest to QCD. The Hamiltonian is 
\begin{equation}
\bm{H}_{\mbox{\tiny SU(3)}} =  \sum_{r, \mu} \sum_{\alpha=1}^8 \left( \bm{E}^\alpha (r,\mu) \right)^2 
 - x \sum_{p \in \mbox{\tiny \{plaquettes\}}} \left(  \bm{Z}(p) + \bm{Z}^\dagger(p) \right)
\label{su3ham}
\end{equation}
where $ \bm{Z}(p) $ was defined in (\ref{generalplaq}). The $ \bm{U} $ operators appearing
in the $ \bm{Z} (p ) $ operators are $ 3 \times 3 $ matrices.  There are eight terms in the 
$ \alpha $ summation in  (\ref{generalu}) corresponding to the eight generators
of SU(3). The commutation relations of the $ \bm{E}^\alpha$ 
with the $ \bm{U} $ obey relations (\ref{generalcom1}) - (\ref{generalcom3}). 

The basic strategy is identical to the SU(2) case, albeit with the added complexity of the SU(3) group. 
Let us again consider what kind of states will be generated by successive applications of the plaquette 
terms in (\ref{su3ham}) on the state satisfying $ \bm{E}^\alpha (r,\mu) | 0 \rangle = 0 $ 
for all $ r$ and $\mu $. Consider again the double plaquette excitation shown in Fig. \ref{fig:su2}. 
We again obtain terms such as $ \bm{U}_{\gamma_L \gamma_R} \bm{U}_{\lambda_L \lambda_R} | 0 \rangle $, 
which may be simplified using the SU(3) version of the identity (\ref{ucombsu2}) \cite{swart63,kogut76}:
\begin{eqnarray}
&& \hspace{-7mm}  U_{\gamma_L \gamma_R }^{\omega'} (\phi) U_{\lambda_L \lambda_R}^{\omega}  (\phi) = \nonumber \\
&&  \hspace{-5mm}  \sum_{\Omega \Gamma_L \Gamma_R} 
\langle \Omega, \Gamma_L | \omega, \lambda_L; \omega', \gamma_L\rangle
\langle \Omega, \Gamma_R | \omega, \lambda_R; \omega', \gamma_R  \rangle 
U_{\Gamma_L \Gamma_R}^{\Omega} (\phi) \nonumber \\
\label{ucombsu3}
\end{eqnarray}
where 
\begin{equation}
U_{\nu \lambda}^{\omega} (\phi) = 
\left[ \exp( i \sum_{\alpha = 1}^8 \tau^\alpha_{\omega} \phi^\alpha ) \right]_{\nu \lambda} 
\label{su3unitary}
\end{equation}
and $ \tau^\alpha_{\omega} $ are the generators of SU(3) for the representation $ \omega $.  The notation in 
(\ref{ucombsu3}) requires some explanation. First recall that every state in SU(3) can be
labeled by five numbers
\begin{equation}
| p, q, T, T^z,Y \rangle,
\label{su3statenotation}
\end{equation}
where $ p $ and $ q $ label the irreducible representation, 
$ T $ and $ T^z $ label the $T$-spin, and $ Y $ labels the hypercharge \cite{carruthers66}. The label $ \omega $ 
 in (\ref{su3unitary}) then corresponds to $ (p,q )$ and the labels $ \lambda_L$ and $ \lambda_R $ 
then label two 
sets of $ (T, T^z, Y )$. For example, the states of the $ \omega = \bm{3} $ representation (i.e. $ p =1 $, $q=0 $) 
are labeled (in the notation of (\ref{su3statenotation})):
\begin{eqnarray*}
| 1 , 0 , \tfrac{1}{2} , \tfrac{1}{2} , \tfrac{1}{3} \rangle & \leftrightarrow & \lambda = 1 \\
| 1 , 0 , \tfrac{1}{2} , -\tfrac{1}{2} , \tfrac{1}{3} \rangle & \leftrightarrow & \lambda= 2 \\
| 1 , 0 , 0 , 0 , -\tfrac{2}{3} \rangle & \leftrightarrow & \lambda= 3 .
\end{eqnarray*}
The states of $ \omega = \bm{3^*} $ are
\begin{eqnarray*}
| 0 , 1 , \tfrac{1}{2} , \tfrac{1}{2} , -\tfrac{1}{3} \rangle & \leftrightarrow & \lambda= 1 \\
| 0 , 1 , \tfrac{1}{2} , -\tfrac{1}{2} , -\tfrac{1}{3} \rangle & \leftrightarrow & \lambda= 2 \\
| 0 , 1 , 0 , 0 , \tfrac{2}{3} \rangle & \leftrightarrow & \lambda= 3 .
\end{eqnarray*}
The $ \langle \Omega, \Gamma |  \omega, \lambda;\omega', \gamma \rangle $ thus denote Clebsch-Gordan coefficients for SU(3). 
We may therefore define
\begin{eqnarray}
& &  |\omega, \lambda_L, \lambda_R \rangle = | p, q ,T_L, T^z_L, Y_L, T_R, T^z_R, Y_R \rangle = \nonumber \\
& & \hspace{30mm} \sqrt{\mbox{dim}(p,q)}
\bm{U}_{\lambda_L \lambda_R}^{\omega} |0 \rangle ,
\label{statesu3}
\end{eqnarray}
where the factor under the square root is the dimension of the irreducible representation $ (p,q) $
\begin{equation}
\mbox{dim}(p,q) = (1+p)(1+q)(1+ \tfrac{1}{2} (p+q)) .
\end{equation}
This factor is necessary 
in order to normalize the states:
\begin{equation}
\langle \omega', \lambda_L', \lambda_R' | \omega, \lambda_L, \lambda_R  \rangle = 
\delta_{\omega \omega'} \delta_{\lambda_L \lambda_L'} \delta_{\lambda_R \lambda_R'} .
\end{equation}
The $ \bm{U}_{\lambda_L \lambda_R}  $ operators appearing in the Hamiltonian (\ref{su3ham}) are all in the 
fundamental representation of SU(3), i.e. the $ \bm{3} $ representation ($ p = 1 $, $ q = 0 $). The 
$ \bm{U}_{\lambda_L \lambda_R}^\dagger  $ operators on the other hand are in the $ \bm{3^*} $ representation 
($ p = 0 $, $ q = 1 $). Terms such as
$ \bm{U}_{\gamma_L \gamma_R}^{\bm{3}} \bm{U}_{\lambda_L \lambda_R}^{\omega} | 0 \rangle $ may then be explicitly 
written 
\begin{widetext}
\begin{eqnarray}
& & \bm{U}_{\gamma_L \gamma_R}^{\bm{3}} | p, q, T_L, T^z_L, Y_L, T_R, T^z_R, Y_R \rangle = \nonumber \\
& & \hspace{10mm} \sum_{(p',q')} \sum_{T_L' = |T_L - t_L|}^ {T_L + t_L} \sum_{T_R' = |T_R - t_R|}^ {T_R + t_R}
\sqrt{\frac{\mbox{dim}(p,q)}{\mbox{dim}(p',q')}} 
\langle p', q', T_L', {T_L^z}', Y_L' |  p ,q , T_L, T^z_L, Y_L; \bm{3} , t_L, t_L^z , y_L \rangle  \nonumber \\
& & \hspace{20mm} 
\times  \langle p', q', T_R', {T_R^z}', Y_R' | p, q, T_R, T^z_R, Y_R; \bm{3}, t_R, t_R^z, y_R  \rangle 
| p', q', T_L',  {T_L^z}', Y_L', T_R', {T_R^z}', Y_R' \rangle ,
\label{su3additionrule}
\end{eqnarray}
\end{widetext}
where the $ (p',q') $ summation is over the three terms $ \{ (p+1,q), (p-1,q+1), (p,q-1) \} $ and the 
$ (t_i, t_i^z, y_i) $ are the corresponding $T$-spin and hypercharge values corresponding to the index 
$ \gamma_i $ ($ i = L,R$).
The hypercharge and the 
$ T^z $-spin simply add together as
\begin{eqnarray}
Y_i' & = & Y_i + y_i \\
{T_i^z}' & = & T_i^z + t_i^z 
\end{eqnarray}
for $ i = L, R $. We also require the rule for operating 
$ \bm{U}_{\gamma_L \gamma_R}^{\bm{3^*}} $ since the analogous
identity to (\ref{su2identity}) does not hold for SU(3). This is given by replacing $ \bm{3} \rightarrow \bm{3^*} $ 
in (\ref{su3additionrule}) and summing over $ \{ (p,q+1), (p+1,q-1), (p-1,q) \} $ in the $ (p',q') $ summation. 

The first term in (\ref{su3ham}) corresponds to operating the Casimir operator for SU(3). We thus 
have \cite{carruthers66}
\begin{equation}
 \sum_\alpha ( \bm{E}^\alpha )^2 | p, q \rangle = \nonumber \tfrac{1}{3} \left( p^2 + q^2 + pq + 3(p+q) \right) | p, q  \rangle , \nonumber  \\
\end{equation}
where we have omitted the $T$-spin and $Y$ labels for brevity. 

We may now write the SU(3) Hamiltonian in a spin language. The basic strategy is the same as for SU(2), where 
we use a modification of the circuit of Bacon, Chuang, and Harrow \cite{bacon04}. In Ref \cite{bacon04} it is 
suggested that {\it qudit} registers are used to keep track of the states for SU($N$). We follow a slightly different
approach, where the same {\it qubit} registers are used. Therefore all operators acting on the 
registers that we discuss in the following may be built from standard SU(2) Pauli matrices. 

For each link associate eight registers 
according to (\ref{statesu3}), which we denote by 
$ | p, q, T_L, T^z_L, Y_L, T_R, T^z_R, Y_R \rangle_{\mbox{\tiny reg}} $. As in the U(1) and SU(2) 
cases we must impose a cutoff on these registers to keep the number of qubits finite. If the highest
representation that is stored is $ ( p_{\mbox{\tiny max}}, q_{\mbox{\tiny max}}) $, then we have registers
in the range $ p = \{0, 1, \dots , p_{\mbox{\tiny max}} \} $ and $ q = \{0,1, \dots , q_{\mbox{\tiny max}} \} $. 
The other registers will then be in the range 
\begin{eqnarray}
T_i& = & \{ 0, \tfrac{1}{2},\dots , \tfrac{1}{2}(p+q) \} , \label{Trange} \\ 
T_i^z & = & \{ -\tfrac{1}{2} (p+q), -\tfrac{1}{2} (p+q) + \tfrac{1}{2},\dots , \tfrac{1}{2}(p+q) \} , \nonumber \\ 
\label{Tzrange}  \\
Y_i & = & \{ -\tfrac{1}{3}(q+2p), -\tfrac{1}{3}(q+2p) + \tfrac{1}{3}, \dots, \tfrac{1}{3}(p+2q) \} \nonumber 
\label{Yrange}, \\ 
\end{eqnarray}
for $ i = L,R $. Define operators that shift the registers as follows
\begin{eqnarray}
P^\pm | p \rangle_{\mbox{\tiny reg}} = | p \pm 1 \rangle_{\mbox{\tiny reg}} & \hspace{10mm} &
Q^\pm | q \rangle_{\mbox{\tiny reg}} = | q \pm 1 \rangle_{\mbox{\tiny reg}} \nonumber \\
T_i^\pm | T_i \rangle_{\mbox{\tiny reg}} = | T_i \pm \tfrac{1}{2} \rangle_{\mbox{\tiny reg}} & \hspace{10mm} &
{T_i^z}^\pm | T_i^z \rangle_{\mbox{\tiny reg}} = | T_i^z \pm \tfrac{1}{2} \rangle_{\mbox{\tiny reg}} \nonumber \\
Y_i^\pm | Y_i \rangle_{\mbox{\tiny reg}} = | Y_i \pm \tfrac{1}{3} \rangle_{\mbox{\tiny reg}}
\end{eqnarray}
for $ i = L,R $. Using (\ref{su3additionrule}) we may write the operator 
corresponding to $ \bm{U}_{\nu \lambda}^{\bm{3}} $:
\begin{eqnarray}
& & \bm{U}_{\nu \lambda}^{\bm{3}} \leftrightarrow  V_{\nu \lambda} =  M_{\nu}^L M_{\lambda}^R \big[  
P^+ {\cal C}_{1 \nu}^L {\cal C}_{1 \lambda}^R   \hat{N}_1 \nonumber \\
& & \hspace{5mm} + P^- Q^+ {\cal C}_{2 \nu}^L {\cal C}_{2 \lambda}^R \hat{N}_2
+ Q^- {\cal C}_{3 \nu}^L {\cal C}_{3 \lambda}^R \hat{N}_3 \big]  ,
\label{su3Voperator}
\end{eqnarray}
where we have defined the operators
\begin{equation}
{\cal C}_{\omega \nu}^i = \left\{
\begin{array}{ll}
T_i^+ \hat{C}_{\omega \nu}^{i(a)}  +  T_i^- \hat{C}_{\omega \nu}^{i(b)} & \nu=1,2 \\
\hat{C}_{\omega \nu}^i & \nu = 3
\end{array} \right. 
\end{equation}
and
\begin{equation}
M_{\nu}^i = \left\{
\begin{array}{ll}
{T_i^z}^+ Y_i^+ & \nu=1 \\
{T_i^z}^- Y_i^+ & \nu=2 \\
(Y_i^-)^2 & \nu=3
\end{array} \right. 
\end{equation}
for $ i = L,R $. The normalization factors are
\begin{equation}
\label{normalizationsu3}
N_\omega = \left\{
\begin{array}{ll}
\sqrt{\frac{\mbox{dim} (p,q)}{\mbox{dim} (p+1,q)}} & \omega=1 \\
\sqrt{\frac{\mbox{dim} (p,q)}{\mbox{dim} (p-1,q+1)}} & \omega=2 \\
\sqrt{\frac{\mbox{dim} (p,q)}{\mbox{dim} (p,q-1)}} & \omega=3 
\end{array} \right.  .
\end{equation}
The $ \hat{C}_{\omega \nu} $ are operators that recover the Clebsch-Gordan coefficients as
defined in Tab. \ref{tab:su3coeffs}. These were calculated using the ``pattern calculus'' of Biedenharn and Louck \cite{biedenharn68}
(see Appendix \ref{sec:app1}). Similarly, the $ \hat{N}_{\omega} $ recover the normalization factors
defined in (\ref{normalizationsu3}). 
The operator corresponding to $ \bm{U}_{\nu \lambda}^{\bm{3^*}} $ may be found using the identity \cite{swart63}
\begin{equation}
\bm{U}_{\nu_L \nu_R}^{\bm{3^*}} = (-1)^{{\cal Q}_L + {\cal Q}_R} (\bm{U}_{-\nu_L  -\nu_R }^{\bm{3}} )^*
\end{equation}
where
\begin{equation}
{\cal Q}_i = T^z_i+ \tfrac{1}{2} Y_i + \tfrac{1}{3}
\end{equation}
and $-\nu_i = (T_i,-T^z_i,-Y_i)$. The phase factor follows from the standard 
choice of phase factors between states of $ \bm{3} $ and $ \bm{3}^* $ \cite{carruthers66}. The desired operator is then
\begin{equation}
\bm{U}_{\nu_L \nu_R}^{\bm{3^*}} \leftrightarrow (-1)^{{\cal Q}_L + {\cal Q}_R} V_{\nu_L \nu_R}^\dagger ,
\label{su3conjVoperator}
\end{equation}
where the $ \dagger $ refers to Hermitian conjugation with respect to the spin operators (e.g.
$ (P^\pm)^\dagger = P^\mp, (T_i^\pm)^\dagger = T_i^\mp, (\hat{C}_{\omega \nu}^i)^\dagger = \hat{C}_{\omega \nu}^i , \dots $). 
This result may be checked by directly constructing the operators in the same way as (\ref{su3Voperator}) with 
the proper $ \bm{3^*} $ Clebsch-Gordan coefficients, and verifying that (\ref{su3conjVoperator}) holds. 

The SU(3) Hamiltonian may be written 
\begin{equation}
H_{\mbox{\tiny SU(3)}} = \sum_{r,\mu} E^2 (r,\mu)
- x \sum_{p \in \mbox{\tiny \{plaquettes\}} }\left( Z(p) + Z^\dagger (p) \right)
\end{equation}
where
\begin{eqnarray*}
&  & Z(p)  =  \sum_{\nu_1 \nu_2 \nu_3 \nu_4} (-1)^{{\cal Q}_1+{\cal Q}_3}  \\
& &V_{\nu_1 \nu_2} (r,\mu)  V_{\nu_2 \nu_3} (r+\mu,\nu)  
V_{\nu_3 \nu_4}^\dagger (r+\nu,\mu)  V_{\nu_4 \nu_1}^\dagger (r,\nu) , \\
\end{eqnarray*}
and 
\begin{equation*}
E^2 | p , q \rangle_{\mbox{\tiny reg}} = \tfrac{1}{3} \left( p^2 + q^2 + pq + 3(p+q) \right) | p, q  \rangle_{\mbox{\tiny reg}} . 
\end{equation*}
This is the final result for SU(3). 

Let us again find the required number of qubits and the dependence of the 
number of operations for a lattice size with $ M $ sites.  The total number of qubits is $ \sim 
8 M D d $, due to the eight registers required for each link each with approximately $ D $ qubits. 
With a similar arrangement of qubits as Fig. \ref{fig:qubitarrangement} we have a linear dependence of the 
number of operations with $ Md $. The analysis for the worst case is identical to the SU(2) case, 
and thus we obtain a worst-case number of operations increasing as $ \propto M^2 d^2 D  $.

\begin{table}
\caption{Clebsch-Gordan coefficients for $ \bm{R} \otimes \bm{3}  $ for SU(3).
The tabulated formulas give the values for 
$ \langle p + \Delta p, q + \Delta q, T + \Delta T, T^z + \Delta T^z, Y + \Delta Y |
 p,q,T,T^z,Y; 1,0, \Delta T, \Delta T^z, \Delta Y \rangle $.  Formulas may be obtained
by combining the results from Table \ref{tab:su2coeffs} and Table \ref{tab:su3isoscalar}, where
$ i = L, R $ is put in as
appropriate.
 \label{tab:su3coeffs}}
\begin{ruledtabular}
\begin{tabular}{ccccccc}
Coefficient & $ \Delta p $ & $ \Delta q $ & $ \Delta T $ & $ \Delta T^z $ & $ \Delta Y $ & Formula \\
\hline
$ C_{11}^{i(a)} $  &  1 & 0 &  1/2 & 1/2  &  1/3 & $ I_{11}^i c_{11}^i $ \\
$ C_{11}^{i(b)} $  &  1 & 0 & -1/2 &  1/2 &  1/3 & $ I_{12}^i c_{21}^i $ \\
$ C_{12}^{i(a)} $  &  1 & 0 &  1/2 & -1/2 &  1/3 & $ I_{11}^i c_{12}^i $ \\
$ C_{12}^{i(b)} $  &  1 & 0 & -1/2 & -1/2 &  1/3 & $ I_{12}^i c_{22}^i $ \\
$ C_{13}^i   $  &  1 & 0 &    0 &    0 & -2/3 & $ I_{13}^i $ \\
$ C_{21}^{i(a)} $  & -1 & 1 &  1/2 &  1/2 &  1/3 & $ I_{21}^i c_{11}^i $ \\
$ C_{21}^{i(b)} $  & -1 & 1 & -1/2 &  1/2 &  1/3 & $ I_{22}^i c_{21}^i $ \\
$ C_{22}^{i(a)} $  & -1 & 1 &  1/2 & -1/2 &  1/3 & $ I_{21}^i c_{12}^i $ \\
$ C_{22}^{i(b)} $  & -1 & 1 & -1/2 & -1/2 &  1/3 & $ I_{22}^i c_{22}^i $ \\
$ C_{23}^i   $  & -1 & 1 &    0 &    0 & -2/3 & $ I_{23}^i $ \\
$ C_{31}^{i(a)} $  &  0 &-1 &  1/2 &  1/2 &  1/3 & $ I_{31}^i c_{11}^i $ \\
$ C_{31}^{i(b)} $  &  0 &-1 & -1/2 &  1/2 &  1/3 & $ I_{32}^i c_{21}^i $ \\
$ C_{32}^{i(a)} $  &  0 &-1 &  1/2 & -1/2 &  1/3 & $ I_{31}^i c_{12}^i $ \\
$ C_{32}^{i(b)} $  &  0 &-1 & -1/2 & -1/2 &  1/3 & $ I_{32}^i c_{22}^i $ \\
$ C_{33}^i   $  &  0 &-1 &    0 &    0 & -2/3 & $ I_{33} $ \\  
\end{tabular}
\end{ruledtabular}
\end{table}

\begin{table}
\caption{Isoscalar factors for $ \bm{R} \otimes \bm{3} $. 
Isoscalar formulas are obtained by substituting 
$ \Omega^\pm_1 = 4p+ 2q \pm 6T_i - 3Y_i +9 \pm3 $, $ \Omega^\pm_2 = 2p - 2q \pm 6T_i + 3Y_i -3 \pm3 $, 
$ \Omega^\pm_3 = 2p+ 4q \pm 6T_i + 3Y_i +3 \pm3 $, $ \Gamma_1 = (1+p)(2+p+q) $, $ \Gamma_2 = (1+p)(1+q) $, 
$ \Gamma_3 = (1+q)(2+p+q) $, $\Upsilon_1 = 432 (1+T_i) $, $\Upsilon_2 = 432T_i $, and $\Upsilon_3 = 36 $ 
into the tabulated formulas. The label $ i = L, R $ is put in as
appropriate. The reduced Wigner operator
used to calculate the formulas are also shown.
\label{tab:su3isoscalar}}
\begin{ruledtabular}
\begin{tabular}{ccc}
Symbol & Reduced Wigner operator& Formula \\
\hline
$ I_{11}^i $  & 
$ \left[
\begin{array}{c}
 1 \\
  1 \ 0  \\
1 \ 0 \ 0\\
1 \ 0 \\
 1  \\ \end{array}
\right] $ & 
$ \sqrt{\frac{\Omega^+_1 (\Omega^+_2+6) (\Omega^+_3+6)}{\Gamma_1 \Upsilon_1}} $
\\
$ I_{12}^i $  &
$ \left[
\begin{array}{c}
 1  \\
 1 \ 0  \\
1 \ 0 \ 0\\
1 \ 0 \\
0 \\ \end{array}
\right] $ &
$ -\sqrt{-\frac{\Omega^-_1 (\Omega^-_2+6) (\Omega^-_3+6)}{\Gamma_1 \Upsilon_2}} $ \\
$ I_{13}^i $  &
$ \left[
\begin{array}{c}
1  \\
1 \ 0 \\
1 \ 0 \ 0\\
0 \ 0 \\
0 \\ \end{array}
\right] $ &
$ \sqrt{\frac{\Omega^-_1 \Omega^+_1}{\Gamma_1 \Upsilon_3}} $ 
\\
$ I_{21}^i $  &
$ \left[
\begin{array}{c}
0 \\
1 \ 0 \\
1 \ 0 \ 0\\
1 \ 0 \\
1\\ \end{array}
\right] $ &
$ - \sqrt{\frac{(6-\Omega^-_1) \Omega^-_2  (\Omega^+_3+6)}{\Gamma_2 \Upsilon_1}} $ \\
$ I_{22}^i $ &
$ \left[
\begin{array}{c}
0 \\
1 \ 0 \\
1 \ 0 \ 0\\
1 \ 0 \\
0 \\ \end{array}
\right] $ &
$ - \sqrt{\frac{(\Omega^+_1-6) \Omega^+_2 (\Omega^-_3+6)}{\Gamma_2 \Upsilon_2}} $ \\
$ I_{23}^i $  &
$ \left[
\begin{array}{c}
 0  \\
1 \ 0 \\
1 \ 0 \ 0\\
0 \ 0 \\
 0  \\ \end{array}
\right] $ &
$ \sqrt{-\frac{\Omega^-_2 \Omega^+_2 }{\Gamma_2 \Upsilon_3}} $  \\
$ I_{31}^i $  &
$ \left[
\begin{array}{c}
0 \\
0 \ 0 \\
1 \ 0 \ 0\\
1 \ 0 \\
1 \\ \end{array}
\right] $ &
$ - \sqrt{\frac{(\Omega^-_1-6) (\Omega^+_2+6) \Omega^-_3}{\Gamma_3 \Upsilon_1}} $  \\
$ I_{32}^i $  &
$ \left[
\begin{array}{c}
0 \\
0 \ 0 \\
1 \ 0 \ 0\\
1 \ 0 \\
0 \\ \end{array}
\right] $ &
$ \sqrt{-\frac{(\Omega^+_1-6) (\Omega^-_2+6) \Omega^+_3}{\Gamma_3 \Upsilon_2}} $  \\
$ I_{33}^i $  &
$ \left[
\begin{array}{c}
0 \\
0 \ 0 \\
1 \ 0 \ 0\\
0 \ 0 \\
 0 \\ \end{array}
\right] $ &
$ \sqrt{\frac{\Omega^+_3 \Omega^-_3}{\Gamma_3 \Upsilon_3}} $ \\  
\end{tabular}
\end{ruledtabular}
\end{table}

\section{\label{sec:init}Preparation of the Initial State}

In this section we discuss how the qubits should be initialized in order to extract observables with respect to 
the low-lying eigenstates of the Hamiltonians. In particular, we have in mind the method of 
extracting the energies of the Hamiltonians as discussed in Refs. \cite{abrams99,somma02}.
In both of these methods, the qubits are prepared in an initial state with a non-zero overlap to the
state of interest, assumed to be the ground state in this case. In the method given by 
Abrams and Lloyd, a sequence of operations in the spirit of phase estimation is performed on the
qubits to extract the eigenvalues of the time-evolution operator $ U(t) $. In the method given by 
Somma {\it et al.}, the expectation value of $ U(t) $ is taken with respect to the prepared state,
then a Fourier transform of measurement is performed on a classical computer to extract eigenvalues. 
The method of operating
the time-evolution operator was discussed at length in Sec. \ref{sec:form}. However, it is 
clear that the initial state preparation must also be done efficiently. 

One potential choice of initial state
is the strong coupling ground state, i.e. $ \bm{E}(r,\mu) | 0 \rangle = 0 $ for all $ r $ and $ \mu $. 
This is a good approximation for 
small values of $ x $, however, for large $ x $ the overlap with the true ground state becomes increasingly 
small. Since it is the weak coupling limit ($ x \rightarrow \infty $) that is generally of interest,
it is clear that a better choice is necessary. In the following, we shall consider how this is done for 
the U(1), SU(2), and SU(3) lattice gauge theories.

\subsection{\label{sec:initu1}Initialization for U(1) Lattice Gauge Theory}

A popular choice of trial state for U(1) lattice gauge theory takes the form \cite{chin84}
\begin{equation}
\bm{\Psi} |0 \rangle =
\prod_{p \in \mbox{\tiny \{plaquettes\}}}  \frac{1}{\sqrt{\cal N}}   \exp \left[ C(x) ( \bm{Z}(p) + \bm{Z}^\dagger(p) ) \right] | 0 \rangle
\label{initialstateu1}
\end{equation}
where the plaquette operator $ \bm{Z}(p) $ was defined in (\ref{u1zp}), $ | 0 \rangle $ is the state defined by
(\ref{u1strongcoup}), and ${\cal N}$  is a suitable normalization factor. $ C(x) $ is a variational parameter to be 
optimized with respect to the coupling $ x $.  This may be performed straightforwardly according to the method given in Ref. \cite{heys85}. 

Since the exponential operator appearing in (\ref{initialstateu1}) is non-unitary, one cannot simply write down the operator that prepares this 
state from the strong coupling ground state $ | 0 \rangle $.  It is still possible to prepare the state given in (\ref{initialstateu1})
by the following procedure. Concentrating on a single plaquette in the lattice, expanding the exponential we obtain
\begin{eqnarray}
\bm{\Psi} (p) =  \frac{1}{\sqrt{\cal N}} \Big[ \zeta_0  + \zeta_1 ( \bm{Z}(p) + \bm{Z}^\dagger(p) ) + \nonumber \\
\zeta_2( \bm{Z}(p)^2 + \bm{Z}^\dagger(p)^2 )
+ \dots \Big] .
\label{initstatezeta}
\end{eqnarray}
For the U(1) case the expansion coefficients may be easily evaluated:
\begin{equation}
\zeta_l = \sum_n \frac{(C(x))^n}{n!} \left( 
\begin{array}{c}
n \\
\tfrac{1}{2}(n-l) \\
\end{array}
\right) = I_l (2C(x)) ,
\end{equation}
where $ I_l $ denotes the modified Bessel function of the first kind. The coefficients $\zeta_l $ are kept until most of the 
dominant terms are found. For $ C(x) = 1$, we find that the most of the dominant terms 
are contained in the range $ 0 \le l \le l_{\mbox{\tiny max}} \approx 5 $.

Now let us see how we may prepare the state (\ref{initstatezeta}) using the formulation given in Sec. \ref{sec:u1}. 
There, we defined operators $ L^\pm $ that increase and decrease the $ | l \rangle $ register by $ \pm 1 $. 
Let us introduce a similar operator acting on a register with the following properties:
\begin{equation}
L(l') |l \rangle_{\mbox{\tiny reg}} = \left\{ 
\begin{array}{ll}
|{l'} \rangle_{\mbox{\tiny reg}} & \hspace{1cm}  \mbox{if } l=0 \\
0 & \hspace{1cm}  \mbox{otherwise} \\
\end{array}
\right. .
\label{restrictedLup}
\end{equation}
The Hermitian conjugate operator to this is 
\begin{equation}
L^\dagger(l')  |l \rangle_{\mbox{\tiny reg}} = \left\{ 
\begin{array}{ll}
|0 \rangle_{\mbox{\tiny reg}} & \hspace{1cm}  \mbox{if } l={l'} \\
0 & \hspace{1cm}  \mbox{otherwise} \\
\end{array}
\right. .
\label{restrictedLdown}
\end{equation}
These operators are restricted versions of the operators $ L^\pm $, that are specialized to 
shifting between the states $ |0 \rangle $ and $ | l \rangle $. 
An example of these operators is given in Appendix \ref{sec:apexample}. We may construct the 
restricted versions of the plaquette operator as follows
\begin{equation}
Z_l (p) = \underbrace{L(l)}_{(r,\mu)} \underbrace{L(l)}_{(r+ \mu,\nu )} \underbrace{L^\dagger(l)}_{(r+\nu,\mu)} 
\underbrace{L^\dagger(l)}_{(r,\nu)} ,
\end{equation}
where we have labeled the links that each of the operators apply underneath each operator. Now consider the 
operator
\begin{equation}
U^{Z(p)}_l  \equiv \exp \left[ \phi (Z_l (p) - Z_l^\dagger (p)) \right] = e^{\left[ -i \phi (i Z_l (p) - iZ_l^\dagger (p) ) \right]} ,
\end{equation}
which is a unitary operator. The advantage of working with these restricted versions of operators is that the 
result of acting these operators may be written in closed form:
\begin{equation}
U^{Z(p)}_l| \square^{l'} \rangle = \left\{
\begin{array}{ll}
\cos \phi | 0 \rangle + \sin \phi | \square^{l} \rangle & \hspace{1cm} \mbox{if } {l'}=0 \\
\cos \phi | \square^l \rangle - \sin \phi | 0 \rangle & \hspace{1cm} \mbox{if } {l'}=l \\
| \square^{l'} \rangle & \hspace{1cm} \mbox{otherwise} \\
\end{array} \right. , 
\label{oneplaqoperator}
\end{equation}
which is easily verified by expanding with respect to $ \phi $. In (\ref{oneplaqoperator}) we defined
\begin{equation}
| \square^l \rangle \equiv Z_l(p) | 0 \rangle_{\mbox{\tiny reg}} .
\end{equation}
The initial state preparation is then performed by the operation
\begin{equation}
\prod_{p \in \mbox{\tiny \{plaquettes\}}} \prod_{l= 1 }^{l_{\mbox{\tiny max}}} U^{Z(p)}_l (\phi_l) U^{Z(p)}_{-l} (\phi_{-l})|0 \rangle_{\mbox{\tiny reg}} .
\label{singleplaqinitspin}
\end{equation}
The values of the $ \phi_l $ are fixed by the $ \zeta_l $ calculated previously. For example, on a single plaquette with 
$ l_{\mbox{\tiny max}} = 1 $, 
\begin{eqnarray}
U^{Z(p)}_{1} (\phi_{1}) U^{Z(p)}_{-1} (\phi_{-1})  |0 \rangle   = 
\cos \phi_{1} \cos \phi_{-1} | 0 \rangle \nonumber \\
+\sin \phi_{-1} | \square^{-1} \rangle  + \sin \phi_{1} \cos \phi_{-1} | \square^1 \rangle ,
\end{eqnarray}
and thus $  \cos \phi_{1} \cos \phi_{-1}  = \zeta_0/ \sqrt{{\cal N}} $, $  \sin \phi_{-1} =  \sin \phi_{1} \cos \phi_{-1} = \zeta_1/ \sqrt{{\cal N}} $. 
This is then repeated for every plaquette in the lattice, which completes the initialization procedure.

\subsection{\label{sec:initsu2}Initialization for SU(2) Lattice Gauge Theory}

A suitable trial wavefunction for SU(2) lattice gauge theory is \cite{heys84}
\begin{equation}
\bm{\Psi} |0 \rangle =
\prod_{p \in \mbox{\tiny \{plaquettes\}}} \frac{1}{\sqrt{{\cal N}}}  \exp \left[ C(x) \bm{Z}(p) \right] | 0 \rangle .
\label{initialstatesu2}
\end{equation}
Concentrating on a particular plaquette, we may again expand the exponential to obtain
\begin{equation}
\bm{\Psi}(p) = \frac{1}{\sqrt{{\cal N}}}  \Big[ \zeta_0  + \zeta_1 \bm{Z}_{1/2} (p)  +  \zeta_2  \bm{Z}_{1}(p)
+ \dots \Big],
\label{initstatezetasu2}
\end{equation}
where
\begin{equation}
\bm{Z}_j(p) =  \mbox{Tr} \left[ \bm{U}^j  (r,\mu) \bm{U}^j (r+ \mu,\nu ) {\bm{U}^j}^\dagger (r+\nu,\mu) {\bm{U}^j}^\dagger (r,\nu) \right]
\label{generalplaqwithj}
\end{equation}
and $ j $ labels the representation of the $ \bm{U} $ operator. The terms in the expansion (\ref{initstatezetasu2}) 
may be calculated with the help of the identity
\begin{equation}
\bm{Z}_{1/2} (p) \bm{Z}_{j} (p) = \bm{Z}_{j-1/2} (p) + \bm{Z}_{j+1/2} (p) .
\end{equation}
In analogy to the U(1) case, let us define restricted versions of the $ J^\pm $ and $ M_{m}^i $ operators as follows:
\begin{eqnarray*}
J(j') |j,m_L,m_R \rangle_{\mbox{\tiny reg}} & = & \left\{
\begin{array}{ll}
|j',m_L,m_R \rangle_{\mbox{\tiny reg}} & \mbox{if } j=0 \\
0 & \mbox{otherwise}
\end{array}
\right. ,\\
M^L (m') |j,m_L,m_R \rangle_{\mbox{\tiny reg}} & = & \left\{
\begin{array}{ll}
|j,m',m_R \rangle _{\mbox{\tiny reg}}& \mbox{if } m_L=0 \\
0 & \mbox{otherwise}
\end{array}
\right. ,\\
M^R (m') |j,m_L,m_R \rangle_{\mbox{\tiny reg}} & = & \left\{
\begin{array}{ll}
|j,m_L,m' \rangle_{\mbox{\tiny reg}} & \mbox{if } m_R=0 \\
0 & \mbox{otherwise}
\end{array}
\right. .
\end{eqnarray*}
The restricted plaquette operator is then defined as
\begin{eqnarray}
& & \hspace{-10mm} Z_{j m_1 m_2 m_3 m_4} (p) = \nonumber \\
& & \underbrace{ J(j) M^L (m_1) M^R (m_2) }_{(r,\mu)} \underbrace{ J(j) M^L (m_2) M^R (m_3) }_{(r+ \mu,\nu )} \nonumber \\
& & \times  \underbrace{ J(j) M^L (m_3) M^R (m_4) }_{(r+\nu,\mu)} \underbrace{ J(j) M^L (m_4) M^R (m_1) }_{(r,\nu)} ,
\end{eqnarray}
where we have again labeled the links that each operator acts on underneath each operator. We now define the unitary 
operator
\begin{equation}
U^{Z(p)}_{j m_1 m_2 m_3 m_4}  \equiv 
\exp \left[ \phi (Z_{j m_1 m_2 m_3 m_4} (p) - Z_{j m_1 m_2 m_3 m_4}^\dagger (p)) \right]  ,
\end{equation}
which has the operation
\begin{eqnarray}
& & \hspace{-4mm} U^{Z(p)}_{j m_1 m_2 m_3 m_4} | \square_{j' m_1' m_2' m_3' m_4'} \rangle   \nonumber \\ 
& &  = \left\{
\begin{array}{ll}
\cos \phi | 0 \rangle + \sin \phi | \square_{j m_1 m_2 m_3 m_4} \rangle &  \mbox{if } j'= m_i'=0 \\
\cos \phi | \square_{j m_1 m_2 m_3 m_4} \rangle - \sin \phi | 0 \rangle &  \mbox{if } j'=j, m_i'=m_i \\
| \square_{j' m_1' m_2' m_3' m_4'} \rangle  & \mbox{otherwise}
\end{array}
\right. , \nonumber \\
\end{eqnarray}
where we defined
\begin{equation}
| \square_{j m_1 m_2 m_3 m_4} \rangle \equiv Z_{j m_1 m_2 m_3 m_4} (p) | 0 \rangle_{\mbox{\tiny reg}} 
\end{equation}
The initial state is then prepared by a product of these operators on the strong coupling ground state:
\begin{equation}
\prod_{p \in \mbox{\tiny \{plaquettes\}}} \prod_{j=1/2}^{j_{\mbox{\tiny max}}}
\prod_{m_1 m_2 m_3 m_4} 
U^{Z(p)}_{j m_1 m_2 m_3 m_4} 
(\phi_{j m_1 m_2 m_3 m_4})
|0 \rangle_{\mbox{\tiny reg}} ,
\label{su2finalinit}
\end{equation}
where each of the $ m_i $ summations run from $ -j_{\mbox{\tiny max}} $ to $ j_{\mbox{\tiny max}} $. 
The $ \phi_{j m_1 m_2 m_3 m_4} $ are fixed using the $ \zeta_j $ coefficients in a similar way to the 
initialization for U(1) lattice gauge theory. This initialization is then carried out on each plaquette 
of the lattice.

\subsection{\label{sec:initsu3}Initialization for SU(3) Lattice Gauge Theory}

The initialization procedure for the SU(3) case is very similar to the U(1) and SU(2) cases. We
will therefore only show a sketch of the results. A suitable trial wavefunction for SU(3) is \cite{chin88}
\begin{equation}
\bm{\Psi} |0 \rangle =
\prod_{p \in \mbox{\tiny \{plaquettes\}}} \frac{1}{\sqrt{{\cal N}}}  \exp \left[ C(x) \left( \bm{Z}(p) + \bm{Z}^\dagger (p) \right) \right] | 0 \rangle .
\label{initialstatesu3}
\end{equation}
In analogy to the U(1) and SU(2) cases, we define restricted plaquette operators $ Z_{\omega \lambda_1 \lambda_2 \lambda_3 \lambda_4} $, 
which creates the state associated with the term 
$ \bm{U}_{\lambda_1 \lambda_2}^\omega  \bm{U}_{\lambda_2 \lambda_3}^\omega 
\bm{U}_{\lambda_3 \lambda_4}^{\omega^\dagger}
\bm{U}_{\lambda_4 \lambda_1}^{\omega^\dagger} $ in the plaquette operator $ \bm{Z}(p) $. The initialization
is then performed by a product of unitary operators defined using the restricted plaquette operators:
\begin{equation}
\prod_{p \in \mbox{\tiny \{plaquettes\}}} \prod_{\omega \lambda_1 \lambda_2 \lambda_3 \lambda_4} 
U^{Z(p)}_{\omega \lambda_1 \lambda_2 \lambda_3 \lambda_4} (\phi_{\omega \lambda_1 \lambda_2 \lambda_3 \lambda_4}) 
|0 \rangle_{\mbox{\tiny reg}} .
\end{equation}

\subsection{\label{sec:initeff}Efficiency of the initialization procedure}

Let us now estimate the dependence of the number of operations of the initialization procedure
with the lattice size $ M$. 
The total number of times the $ U^{Z(p)} (\phi) $ operator must be applied is clearly proportional to the number of 
plaquettes in the lattice $ \propto d M  $. Therefore, with a layout of qubits that preserves the local nature of the 
operation, the number of operations will be $ \propto M d $. However, an unfavorable layout of qubits can increase 
the number of operations
required for each $ U^{Z(p)} (\phi) $ by a factor of the number of qubits used, which is equal to $ D d M $. At worst, 
we therefore find that the total number of operations is proportional to $ D d^2 M^2 $.

\subsection{\label{sec:initother}Other choice of initialization}

Our analysis in the previous subsections were restricted to considering particular forms of trial 
wavefunctions. These are clearly not the only trial wavefunctions that may be used, for example, 
correlated plaquette terms may also be included \cite{hamer00}. These 
trial wavefunctions may also be constructed using the methods given in this section. Inclusion of 
states that have long-distance correlations over the lattice will naturally increase the 
complexity of the initialization procedure by a factor proportional to the size (e.g. the length of the 
perimeter 
of the Wilson loop operator) of the state being created. Since this is bounded by the size of the lattice,
the total number of operations will still be $ \sim \mbox{poly} (M) $.

\section{\label{sec:meas} Measurements}

To obtain results from the simulation, one must be able to extract quantities such as the 
expectation value of operators and the spectrum of the Hamiltonian operator. The energy
spectrum of the Hamiltonian may be obtained following the method outlined in the beginning of Sec. \ref{sec:init}.
The most interesting expectation values to be evaluated are of the Wilson loop operators, defined as
\begin{equation}
\bm{W}_\Gamma = \mbox{Tr} \prod_{r,\mu \in \Gamma } \bm{U} (r,\mu) ,
\end{equation}
where $ \Gamma $ denotes the contour in the lattice that the Wilson loop follows, and $ \bm{U} (r,\mu) $
denotes the unitary operator defined for the link labeled by $ (r,\mu) $ for SU($N$). We may immediately 
transcribe these operators into those using spin operators using the results of Sec. \ref{sec:form}.
For example, for SU(2), 
\begin{eqnarray}
W_\Gamma & = & \sum_{m_1,m_2,\dots,m_N} V_{m_1 m_2} (r_1,\mu_1) V_{m_2 m_3} (r_2,\mu_2)  \dots \nonumber \\
& & \hspace{10mm} \times V_{m_N m_1} (r_N,\mu_N) .
\end{eqnarray}
Analogous results follow for the U(1) and SU(3) cases hold. Expectation values of this operator may be 
found for example by the general phase estimation method \cite{ortiz01}. This involves performing the operation
\begin{equation}
U(t) = e^{-i W_\Gamma} ,
\end{equation}
which may be performed in a similar way to the methods explained in Sec. \ref{sec:form}.

\section{\label{sec:conc}Summary and Conclusions}

We have reformulated the U(1), SU(2), and SU(3) lattice gauge theories into a form that can be simulated
using spin qubit manipulations on a universal quantum computer. This was done by constructing spin operator
versions of the link operators $ \bm{E} $ and $ \bm{U} $ appearing in the Hamiltonian versions of the 
theories. Each link on the lattice is represented
by a set of registers that keep track of the state of the link. In all cases, a cutoff corresponding to the 
maximum representation of the group was found to be necessary to keep the total number of qubits
required for the simulation finite. This is necessary as even a single link has an infinite dimensional
Hilbert space. These high representation states have an energy that grows as the square of the index that 
labels the representation (e.g. $ j(j+1)$ for SU(2)), and thus the impact of truncating these states
can be carefully controlled, as has been found in past studies. 
For the SU(2) and SU(3) cases, a calculation of Clebsch-Gordan coefficients is required for 
every application of the 
plaquette operator $ \bm{Z} (p ) $.  Explicit formulas were given for the Clebsch-Gordan coefficients for 
SU(2) and SU(3), which are precalculated on a classical computer for a given $ j_{\mbox{\tiny max}} $ or 
$ p_{\mbox{\tiny max}},q_{\mbox{\tiny max}} $ respectively. 
This is then written in operator form which 
requires an overhead corresponding to $ \sim \mbox{poly} (j_{\mbox{\tiny max}}) $ or 
$ \sim \mbox{poly} (p_{\mbox{\tiny max}},q_{\mbox{\tiny max}}) $ respectively during the simulation. 
The number of qubits necessary to keep track of a lattice with $ M $ lattice sites and dimension $d$ 
was found to be $ \sim d D M $, where $ D $ is the number of qubits required to keep track of a single link state. 
The number of operations required to perform the Hamiltonian evolution of the system
was found to be proportional to between a linear and quadratic function of the total number of lattice sites. 
This is dependent on the architecture of the quantum computer that is available. The simulation may therefore
be said to be efficiently implementable in this sense. 
 Other aspects of the simulation, such as the initialization of the qubits into a suitable configuration 
for extracting low-energy properties of the system were also considered and were found to have a similar 
dependence with lattice size, and therefore does not spoil the efficiency of the whole procedure. 

With the current technology available, the kind of simulation that we suggest in this paper 
is clearly still a long way off. For example, a typical size of lattice for a QCD simulation today has 
the order of $ 10^4 $ lattice sites. By our scheme for SU(3), we would require $ \sim 10^5 D $ qubits, 
where $ D $ is the size of the register dependent on the cutoff that is employed. 
Note that this does not include 
any quantum error correction, and therefore the real figure will be larger than this. However, the 
method here has the advantage that it is virtually an {\it exact} calculation, up to the cutoff on 
each link which is controllable such that its impact is negligible. All the states associated with the 
low-lying excitations are included. With the inclusion of the fermion fields, 
which adds a further $ \sim M $ qubits to the total, this would allow a direct and exact simulation of QCD 
itself -- an unthinkable prospect on a classical computer. Although significant advances have been 
made in the lattice QCD programme in the past few years \cite{davies04}, such an ``exact'' 
simulation may be a useful check on the results of the classical simulations and may even 
offer new insights into the physics of QCD and related models. 

We have restricted our focus in this paper to pure gauge Hamiltonians, i.e. no fermions. 
For a realistic simulation of QCD, we also require fermions in addition to the gauge degrees of freedom 
that were examined in this paper. One way that fermions may be included into the formulation is to 
transform the fermion operators to spin operators via a Jordan-Wigner transformation
 \cite{ortiz01}.  This involves long strings of operators that are of order the size of the lattice $ M $. 
The space required for including the fermion 
adds $ M $ more qubits to the entire lattice, which far less than the number required for the gauge fields.
This should therefore not affect the main results of this work. 

We note that the same methods as that given here may be straightforwardly generalized to SU($N$) 
lattice gauge 
theories, and we expect that these may also be efficiently implemented. 
One vital ingredient that makes this 
possible is due to the Clebsch-Gordan coefficients for the product of the 
fundamental representation of the 
group to an arbitrary representation of the group according to the ``pattern calculus'' of 
Biedenharn and Louck \cite{biedenharn68}.

There are many ways that the implementation that we describe in this paper 
may be improved, particularly with respect to the total number of qubits. 
In particular, there are many wasted states on a particular link associated with 
states that do not exist. For example, for the SU(2) case, unphysical states such as 
$ | j, m_L>j, m_R>j \rangle_{\mbox{\tiny reg}} $ are included, as far as the 
qubit implementation is concerned. Likewise, unphysical gauge variant states are also included. 
We chose to ignore such inefficiencies as this would
introduce further complexities that are unnecessary for the main goals of this paper,
although these should be possible, if necessary.

\begin{acknowledgments}
T. B. acknowledges the support of a JSPS fellowship and 
wishes to thank Cyrus Master for helpful discussions. 
\end{acknowledgments}

\appendix

\section{\label{sec:apexample}Some Simple Register Operators}

To illustrate the methods used in Sec. \ref{sec:form}, 
let us consider a simple example of the operators that could be used to manipulate the registers. 
Considering the U(1) Hamiltonian (\ref{u1spinhamil}), we see that this consists of
operators $ L^\pm $ and $ E^2 $ that are defined according to (\ref{e2operatoru1})-(\ref{voperatoru1}). 
A simple, but qubit 
inefficient way that this could be implemented is as follows
\begin{eqnarray}
L^+ = \sum_{l=-l_{\mbox{\tiny max}}}^{l_{\mbox{\tiny max}}-1} \sigma^-_l \sigma^+_{l+1} \\
L^- = \sum_{l=-l_{\mbox{\tiny max}}}^{l_{\mbox{\tiny max}}-1} \sigma^+_l \sigma^-_{l+1} \\
E^2 =  \sum_{l=-l_{\mbox{\tiny max}}}^{l_{\mbox{\tiny max}}} l^2 (\sigma^z_l + 1 )/2 ,
\end{eqnarray}
where the $ \sigma^\pm $ and $ \sigma^z $ are Pauli spin operators. 
These operators are designed to act on states with a single up-spin: $ | \downarrow \downarrow 
\dots \uparrow \dots \downarrow \rangle $. The location of the up-spin keeps track of the 
register:
\begin{equation}
| l \rangle_{\mbox{\tiny reg}} = \sigma_l^+ | \downarrow \downarrow \dots \downarrow \rangle .
\end{equation}
A similar type of register was considered in Ref. \cite{somma03}. Similar operators are used 
in the SU(2) and SU(3) Hamiltonians. 

The operators that recover the Clebsch-Gordan coefficients may be constructed by a linear combination of 
diagonal operators that act on the registers. Considering the SU(2) case, let us construct the operator
associated with the coefficients $ \hat{c}_{mn} $. We may write
\begin{equation}
\label{cgcoeffoperator}
\hat{c}_{mn} = a_0 + a_1 J^z + a_2 M^z + a_3 (J^z)^2 + \dots,
\end{equation}
where
\begin{eqnarray*}
J^z |j \rangle_{\mbox{\tiny reg}} = j |j \rangle_{\mbox{\tiny reg}} \\
M^z |m \rangle_{\mbox{\tiny reg}} = m |m \rangle_{\mbox{\tiny reg}} 
\end{eqnarray*}
and the $ a_i $ are coefficients to be determined. The number of terms in (\ref{cgcoeffoperator}) is 
determined by the number of distinct $ (j,m) $ combinations. For a $ j_{\mbox{\tiny max}} = 1 $, there 
are three $ (j,m) $ states, and thus there are three terms in (\ref{cgcoeffoperator}). In this case, the coefficients for
$ \hat{c}_{11} $ will be $ a_0 = 1 $, $ a_1 = 1+1/\sqrt{2} $, and $ a_2 = 1- 1/\sqrt{2} $.

Restricted versions of the operators considered in Sec. \ref{sec:init} (i.e. Eqns. (\ref{restrictedLup})
and (\ref{restrictedLdown})) may be constructed 
as follows:
\begin{eqnarray}
L (l) & = & \sigma^-_0  \sigma^+_l  \\
L^\dagger(l) & = & \sigma^+_0  \sigma^-_l  .
\end{eqnarray}

\section{\label{sec:app1}Calculation of Clebsch-Gordan Coefficient Formulas}

Clebsch-Gordan coefficients in Tables \ref{tab:su2coeffs} - \ref{tab:su3isoscalar} are 
calculated using the ``pattern calculus'' method developed by Biedenharn and Louck 
\cite{biedenharn68,louck70}. The method allows the calculation of Clebsch-Gordan coefficients
for the combination of an arbitrary state of SU($N$) with the fundamental representation of the 
group. For example, Clebsch-Gordan coefficients for $ j \otimes \tfrac{1}{2} $ in SU(2), or 
$ \bm{R} \otimes \bm{3}  $ in SU(3) may be found. 

Each finite dimensional irreducible representation is specified by a set of $ N $ ordered 
integers (positive, zero, or negative):
\begin{equation}
[m]_N = [m_{1N} m_{2N} \dots m_{NN}]
\end{equation}
where $ m_{1N} \ge m_{2N} \ge \dots \ge m_{NN} $. A state in this irreducible representation 
is represented by a triangular array called a Gelfand pattern:
\begin{widetext}
\begin{equation}
\left| 
\left(
\begin{array}{ccccccc}
m_{1N} &           & m_{2N}  &  \dots  & m_{N-1 N} &            & m_{NN} \\
       & m_{1 N-1} &         &  \dots  &           & m_{N-1 N-1}&        \\
       &  \ddots   &         &         &           & \iddots    &        \\
       &           & m_{12}  &         & m_{22}    &            &        \\
       &           &         & m_{11}  &           &            &        \\      
\end{array}
\right)
\right> .
\label{gelfandpat}
\end{equation}
\end{widetext}
The integers $ m_{ij} $ satisfy the ``betweenness'' conditions
\begin{equation}
m_{i j+1} \ge m_{ij} \ge m_{i+1 j+1}
\label{betweenness}
\end{equation}
The corresponding state in the $ | J, M \rangle $ or $ | p,q,T,T^z,Y \rangle $ notation 
for a particular Gelfand pattern may be found by the following 
correspondences -- for SU(2) \cite{louck70}
\begin{eqnarray}
J & = & \tfrac{1}{2}(m_{12}-m_{22}), \\
M & = & m_{11} - \tfrac{1}{2}(m_{12} + m_{22}),
\end{eqnarray}
For SU(3) we use the relations
\begin{eqnarray}
p & = & m_{13} - m_{23}, \\
q & = & m_{23} - m_{33}, \\
T & = & \tfrac{1}{2}(m_{12} - m_{22}) ,\\
T^z & = & m_{11} - \tfrac{1}{2}(m_{12} + m_{22}), \\
Y & = & m_{12} + m_{22} - \tfrac{2}{3} ( m_{13} + m_{23} + m_{33} ) .
\end{eqnarray}
Using the above relations, the method given in Sec. III of Ref. \cite{biedenharn68} may be used to 
generate the matrix elements of reduced Wigner operators. For SU(2), matrix elements of the 
reduced Wigner operators are equal to the Clebsch-Gordan coefficients for SU(2). For SU(3), 
the matrix elements of the reduced Wigner operators are the isoscalar factors for SU(3). The 
formulas can be calculated up to a $ \pm $ sign. The sign is fixed by comparison to tables of 
Clebsch-Gordan and isoscalar factors (see for example Ref. \cite{kaeding95}).

\bibliography{paper}

\end{document}